\documentclass[sigconf]{acmart}

\AtBeginDocument{%
  }

\setcopyright{acmlicensed}
\copyrightyear{2025}
\acmYear{2025}
\setcopyright{cc}
\setcctype{by-nc-sa}
\acmConference[CIKM '25]{Proceedings of the 34h ACM International Conference on Information and Knowledge Management}{November 10--14, 2025}{Seoul, Republic of Korea}
\acmBooktitle{Proceedings of the 34th ACM International Conference on Information and Knowledge Management (CIKM '25), November 10--14, 2025, Seoul, Republic of Korea}
\acmDOI{10.1145/3746252.3760968}
\acmISBN{978-8-4007-2040-6/2025/11}

\usepackage[table,xcdraw]{xcolor}
\usepackage{multirow}
\usepackage{balance}

\begin{document}

\title{Query, Decompose, Compress: Structured Query Expansion for Efficient Multi-Hop Retrieval}

\author{Jungmin Yun}
\orcid{0000-0001-6868-286X}
\affiliation{%
  \institution{Chung-Ang University}
  \city{Seoul}
  \country{Republic of Korea}
}
\email{cocoro357@cau.ac.kr}

\author{Youngbin Kim}
\orcid{0000-0002-2114-0120}
\authornote{Corresponding author}
\affiliation{%
  \institution{Chung-Ang University}
  \city{Seoul}
  \country{Republic of Korea}
}
\email{ybkim85@cau.ac.kr}

\renewcommand{\shortauthors}{Jungmin Yun and Youngbin Kim}

\begin{abstract}
  Large Language Models (LLMs) have been increasingly employed for query expansion. However, their generative nature often undermines performance on complex multi-hop retrieval tasks by introducing irrelevant or noisy information. To address this challenge, we propose DeCoR (Decompose and Compress for Retrieval), a framework grounded in structured information refinement. Rather than generating additional content, DeCoR strategically restructures the query's underlying reasoning process and distills supporting evidence from retrieved documents. It consists of two core components tailored to the challenges of multi-hop retrieval: (1) Query Decomposition, which decomposes a complex query into explicit reasoning steps, and (2) Query-aware Document Compression, which synthesizes dispersed evidence from candidate documents into a concise summary relevant to the query. This structured design ensures that the final query representation remains both robust and comprehensive. Experimental results demonstrate that, despite utilizing a relatively small LLM, DeCoR outperforms strong baselines that rely on larger models. This finding underscores that, in complex retrieval scenarios, sophisticatedly leveraging the reasoning and summarization capabilities of LLMs offers a more efficient and effective solution than relying solely on their generative capability.
\end{abstract}



\maketitle

\section{Introduction}
Information Retrieval (IR) systems aim to retrieve relevant information from large corpora in response to a user's query. A primary challenge, however, is that queries are often brief or omit essential terms and concepts, preventing them from fully capturing the complexity of the underlying information need~\cite{lin-etal-2023-decomposing, 10.1145/3534965, 10.1145/3488560.3498440}. Query expansion addresses this limitation by enriching the original query with supplementary information and semantically related concepts~\cite{sharma2023query, liu2022query, rocchio1971relevance, zheng2020bert}. This process constructs a more comprehensive representation of user intent, thereby enabling broader and more accurate document retrieval and ultimately enhancing the overall effectiveness of IR systems~\cite{10.1145/3624988, sharma2022query, nogueira2019document}.

Recent advances in query expansion increasingly leverage the generative capabilities of LLMs to enrich query semantics~\cite{jagerman2023query, jia-etal-2024-mill, mackie2023generative, he2022rethinking}. Query2Doc~\cite{wang-etal-2023-query2doc} generates a pseudo-reference appended to the original query to guide retrieval, while HyDE~\cite{gao-etal-2023-precise} synthesizes a hypothetical document that directly serves as the basis for retrieval. Despite these advances, generative query expansion methods face notable challenges. While expanded content can be informative, it often contains irrelevant or noisy information, which dilutes relevance signals and impairs retrieval performance~\cite{weller-etal-2024-generative}. Furthermore, their effectiveness depends heavily on the quality of the underlying LLM, as less capable models may generate hallucinated or distracting content that undermines retrieval accuracy.

Another critical limitation of existing query expansion research is its predominant focus on single-hop queries. In contrast, multi-hop queries require integrating and reasoning over evidence scattered across multiple documents, which presents a distinct challenge~\cite{trivedi2022interleaving, ho2020constructing}. A major difficulty is that the initial query often omits explicit intermediate reasoning steps, making it insufficient to capture the dependencies and reasoning chains necessary for effective retrieval in complex scenarios~\cite{schnitzler2024morehopqa, geva2021did, trivedi-etal-2022-musique}.

To address these issues, we propose \textbf{DeCoR} (\textbf{De}compose and \textbf{Co}mpress for \textbf{R}etrieval), a novel query expansion framework tailored for multi-hop retrieval. DeCoR shifts the focus from noisy content generation to structured information refinement. Rather than injecting pseudo-information from synthetic content, it maximizes the utility of retrieved documents by explicitly restructuring the query's reasoning process and selectively distilling relevant evidence from the retrieval context.

DeCoR operates in two key stages. First, it incorporates \textbf{Query Decomposition} into the expansion process, explicitly modeling multi-step reasoning paths to enhance interpretability and increase query diversity. Second, following candidate document retrieval, a \textbf{Query-aware Document Compression} module condenses dispersed evidence from documents into concise, query-relevant representations. Despite using a relatively small LLM, DeCoR achieves superior performance compared to prominent baselines that rely on substantially larger models.

In summary, our core contribution lies in introducing DeCoR as a new paradigm for query expansion that enhances both efficiency and robustness in complex IR. By strategically reasoning over, structuring, and refining existing information, DeCoR demonstrates that effective retrieval can be achieved through principled refinements rather than through the sheer capacity of large-scale models. The main contributions of this paper are as follows:
 \begin{itemize}
     \item \textit{Systematic Analysis of Multi-Hop Queries.} We present the first systematic study of query expansion methods in multi-hop retrieval, showing that existing approaches often degrade performance by introducing irrelevant information.
     \item \textit{Structured Information Refinement.} We propose a framework that refines existing information rather than generating new content, combining Query Decomposition with Query-aware Document Compression to improve retrieval relevance.
     \item \textit{Efficiency with Strong Performance.} Despite using a relatively small LLM, DeCoR consistently outperforms much larger models, demonstrating that smaller models can be both effective and efficient for complex retrieval tasks.
 \end{itemize}

\section{Methodology}

\begin{figure}[t!]
  \centering
  \includegraphics[width=1.0\columnwidth]{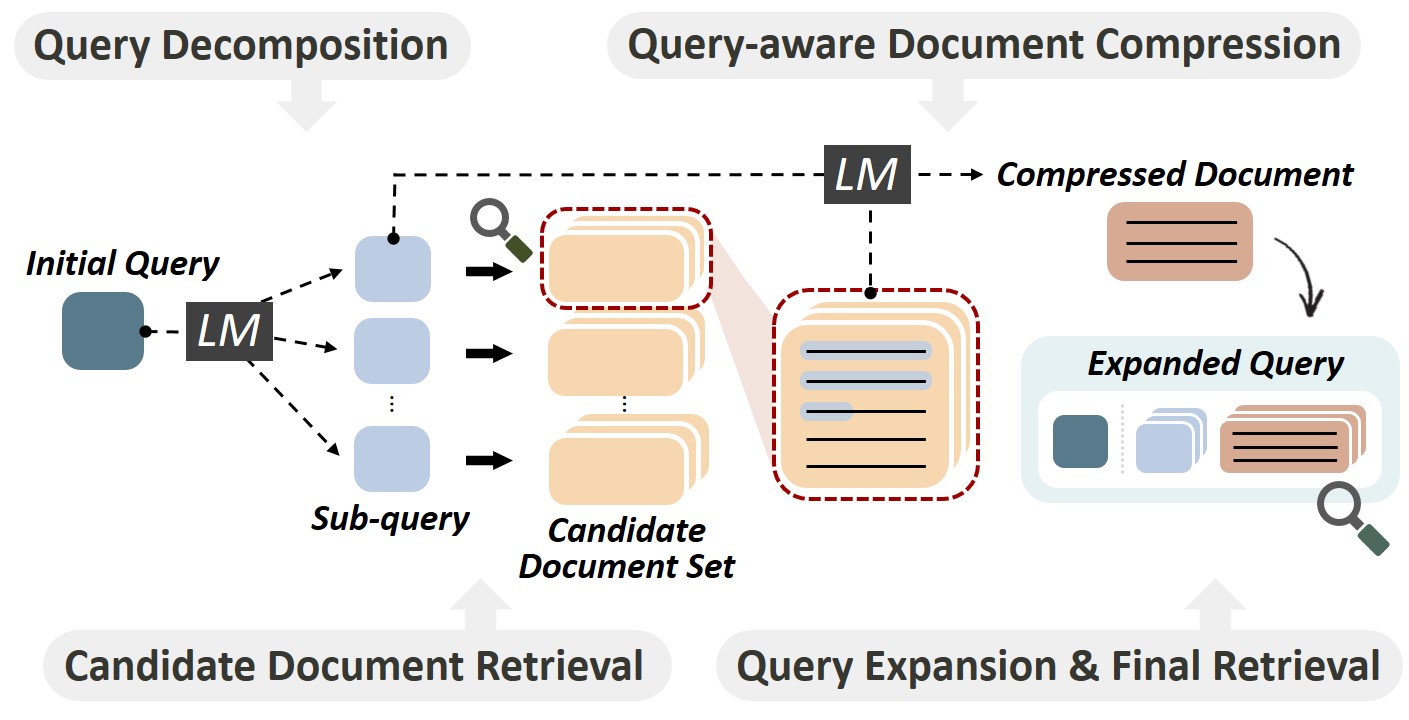}
  \caption{Overall pipeline of DeCoR.}
  \Description{Figure illustrating the overall pipeline of the proposed DeCoR framework.}
  \label{fig1}
\end{figure}

\subsection{Problem Formulation}
We define the task of IR as follows: Given an initial query $q$ and a document collection $C$, a retriever model $M$ retrieves a set of top-$k$ documents $D=\{d_{1}, \ldots, d_{k}\}$, where $D \subseteq C$. The goal of query expansion is to transform the initial query $q$ into a reformulated query $q'$, such that $q'$ preserves the original information need while providing additional contextual signals that facilitate the retrieval of more relevant documents. Formally, when the same retriever $M$ is applied to $q'$, it produces a new document set $D'$, which is expected to exhibit a higher degree of relevance to the initial query $q$ than the original set $D$. This formulation underscores the central challenge of query expansion: enriching queries in a manner that improves retrieval effectiveness without altering or distorting the user’s underlying intent.

\begin{table}[t!]
\caption{Task-specific Prompts Design in DeCoR.}
\label{tab:prompt}
\resizebox{0.9\columnwidth}{!}{
\begin{tabular}{p{9cm}}
\toprule
\rowcolor[HTML]{EFEFEF} 
\textbf{Prompt for Query Decomposition} \\ 
You are a helpful assistant that breaks down complex, multi-hop questions into a list of simpler, independent sub-queries. Each sub-query should reflect a single reasoning step and be answerable on its own. \\
If the question is already simple, return a Python-style list with just the original question. \\
\\
Examples: \\
Question: When was the creator of The Painter's Studio born? \\
Sub-queries: ["Who created The Painter\'s Studio?", "When was the creator of The Painter\'s Studio born?"]\\ \\
Question: What is the capital of Korea?\\ 
Sub-queries: ["What is the capital of Korea?"]\\
\midrule
\rowcolor[HTML]{EFEFEF} 
\textbf{Prompt for Query-aware Document Compression} \\  
You are a helpful assistant that concisely summarizes only the key information from the source documents that is relevant to answering the question. \\
Exclude unrelated content and avoid using pronouns.
\\
\bottomrule
\end{tabular}
}
\end{table}

\subsection{DeCoR Framework}
The overall pipeline of DeCoR is illustrated in Figure~\ref{fig1}. The prompts used for each component, including Query Decomposition and Query-aware Document Compression, are presented in Table~\ref{tab:prompt}.

\subsubsection{Query Decomposition} We employ a query decomposition approach, leveraging LLMs, in which the initial query $q$ is divided into a set of sub-queries $\{q_{1}^{sub}, \ldots, q_{m}^{sub}\}$, where $m$ denotes the number of sub-queries derived from $q$. This strategy serves several key purposes in enhancing the effectiveness of query expansion. First, it breaks down a complex information need into more fine-grained and manageable units, enabling the retriever to address each component with greater precision. Second, it facilitates the exploration of diverse facets and perspectives of the original query, as each sub-query potentially targets a different aspect of the information need, thereby producing more comprehensive retrieval results. Finally, in scenarios requiring multi-step reasoning, sub-queries can be explicitly aligned with distinct logical steps in the reasoning chain. By systematically addressing these sub-components, Query Decomposition mitigates the limitations of treating the query as a single monolithic unit and enables DeCoR to more effectively handle reasoning-intensive tasks, ultimately leading to more accurate and robust retrieval.

\subsubsection{Candidate Document Retrieval} For each sub-query $q_j^{sub} \in q$, we perform an individual retrieval step to obtain a set of candidate documents $D_j^{cand}=\{d^{cand}_{j,1}, \ldots, d_{j,n}^{cand}\}$, where $n$ denotes the number of candidate documents retrieved for that sub-query. Because the single initial query $q$ is expanded into $m$ sub-queries, the retrieval stage inevitably incurs additional computational cost. To ensure that this candidate document retrieval process is both fast and efficient, we adopt BM25~\cite{robertson1994some}, a well-established and effective sparse retrieval algorithm, as the initial retriever for each sub-query.

\subsubsection{Query-aware Document Compression} The candidate document set $D^{cand}$ obtained from the initial retrieval often contains lengthy documents, many of which include content irrelevant to specific sub-queries. Such irrelevant portions can act as informational noise in subsequent stages, potentially obscuring relevance signals or leading to the retrieval of non-relevant documents. To mitigate this issue and improve the efficiency of document representations, we employ a query-aware document compression approach.

Given the candidate set $D_j^{cand}$ associated with a sub-query $q_j^{sub}$, we adopt a \textit{concatenate-then-compress} strategy, which has been empirically shown to outperform document-wise compression. Specifically, all documents in $D_j^{cand}$ are first concatenated into a single context sequence, which is then passed to the LLM to generate a compressed document $d_j^{comp}$. The compression is guided by three complementary mechanisms: (i) \textit{global salience detection}, where the LLM extracts salient information across multiple documents and prioritizes content most relevant to the sub-query; (ii) \textit{cross-document evidence integration}, where complementary information dispersed across different sources is merged into more complete and coherent representations, enriching the context; (iii) \textit{semantic deduplication}, where redundant or overlapping expressions are eliminated through semantic reasoning, yielding concise and information-dense outputs. By reconstructing documents around a query-specific informational core, this approach ensures that subsequent retrieval stages operate only on information aligned with the intent of each sub-query, thereby providing a solid foundation for improving both accuracy and efficiency in multi-hop retrieval tasks.

\subsubsection{Query Expansion \& Final Retrieval} 
The final retrieval is performed using an expanded query representation that integrates the initial query with its sub-queries and their corresponding compressed documents. This design enriches the query with structured contextual information, thereby enhancing retrieval performance. To address input length constraints and construct an effective representation, query expansion is conducted in the feature space by averaging embeddings. Formally, given an initial query $q$ and a set of $m$ sub-query-compressed document pairs ${(q_i^{sub},d_i^{comp})}^m_{i=1}$, the expanded query embedding $e_{exp}$ is defined as:
\begin{equation}
    e_{exp}=\frac{1}{m+1}(E(q)+\sum_{i=1}^{m}E([q_i^{sub};d_i^{comp}])),
\end{equation}
\\where $E(\cdot)$ denotes the encoder that maps text into dense embeddings, and [;] indicates concatenation. This ensures that each sub-query and its associated document contributes equally to the final representation, resulting in a balanced and robust embedding.

The expanded query embedding $e_{exp}$ is used for dense retrieval. The relevance of each document in the pre-indexed collection, represented by its embedding $e_{doc}$, is computed using cosine similarity:
\begin{equation}
    Rel(e_{exp}, e_{doc})=\frac{e_{exp} \cdot e_{doc}}{ ||e_{exp}||  ||e_{doc}||}.
\end{equation}
\\Finally, all candidate documents are ranked in descending order of their relevance scores to generate the final retrieval results.

\section{Experiments}

\subsection{Experimental Setup}

Evaluating IR performance on multi-hop queries is particularly challenging due to the scarcity of dedicated benchmarks. To address this limitation, we adopt the MultiHop-RAG dataset~\cite{tangmultihop} for evaluation. Although originally developed for multi-hop question answering, its primary challenge lies in retrieving and reasoning over multiple documents, making it a rigorous and suitable benchmark for assessing the effectiveness of our proposed DeCoR in multi-hop retrieval tasks. The dataset consists of 2556 queries, with supporting evidence for each query distributed across two to four documents.

For comparison, we include strong generative query expansion baselines such as HyDE~\cite{gao-etal-2023-precise} and Query2Doc~\cite{wang-etal-2023-query2doc}. Our experiments further incorporate multiple dense retrievers and embedding models of varying parameter sizes, including \texttt{Contriever}~\cite{izacardunsupervised}, \texttt{e5-base-v2}~\cite{wang2022text}, and \texttt{bge-large-en-v1.5}~\cite{bge_embedding}. For DeCoR, we retrieve the top-$5$ candidate documents ($n=5$) during the initial retrieval stage. Query Decomposition and Query-aware Document Compression are implemented using the instruction-tuned \texttt{Qwen2.5-7B}, with vLLM employed for efficient inference.

\subsection{Experimental Results}

\begin{table}[t!]
\caption{Experimental results on IR. The best performance is highlighted in \textbf{bold}.}
\label{Tab:1}
\resizebox{0.9\columnwidth}{!}{
\begin{tabular}{lcccc}
\toprule
                           & \textbf{Hits@10} & \textbf{Hits@4} & \textbf{MAP@10} & \textbf{MARR@10} \\ \hline\hline
\rowcolor[HTML]{EFEFEF} 
\texttt{Contriever}        & 62.75            & 48.43           & 17.98           & 40.57            \\
+ HyDE                     & 60.44            & 44.97           & 17.01           & 37.38            \\
+ Query2Doc                & 62.31            & 47.58           & 17.57           & 38.50            \\
+ DeCoR (\textit{ours})                     & \textbf{64.48}   & \textbf{50.91}  & \textbf{20.07}  & \textbf{44.60}   \\ \hline
\rowcolor[HTML]{EFEFEF} 
\texttt{e5-base-v2}        & 69.05            & 53.61           & 19.60           & 44.55            \\
+ HyDE                     & 67.85            & 53.08           & 19.54           & 44.73            \\
+ Query2Doc                & 68.96            & 53.44           & 19.80           & 45.25            \\
+ DeCoR (\textit{ours})                     & \textbf{72.42}   & \textbf{59.42}  & \textbf{22.66}  & \textbf{51.95}   \\ \hline
\rowcolor[HTML]{EFEFEF} 
\texttt{bge-large-en-v1.5} & 68.96            & 54.63           & 19.97           & 45.20            \\
+ HyDE                     & 66.74            & 50.73           & 19.10           & 42.79            \\
+ Query2Doc                & 67.58            & 51.49           & 19.51           & 44.19            \\
+ DeCoR (\textit{ours})                     & \textbf{72.06}   & \textbf{58.23}  & \textbf{22.70}  & \textbf{51.39}   \\ \bottomrule
\end{tabular}
}
\end{table}

\subsubsection{Main Results}
To evaluate the effectiveness of our proposed DeCoR, we adopt Hits@10, Hits@4, MAP@10, and MARR@10 as the primary evaluation metrics for IR. Table~\ref{Tab:1} presents experimental results across three base retrievers—\texttt{Contriever}, \texttt{e5-base-v2}, and \texttt{bge-large-en-v1.5}—combined with different query expansion strategies. Notably, existing query expansion techniques such as HyDE and Query2Doc frequently result in performance degradation across all retrievers. This suggests that their expanded queries often introduce irrelevant or noisy content, weakening relevance signals and hindering effective retrieval.

In contrast, DeCoR consistently improves retrieval performance across all retrievers and metrics. For instance, with \texttt{e5-base-v2}, DeCoR achieves the highest scores on all metrics, including Hits@10 (72.42), Hits@4 (59.42), MAP@10 (22.66), and MARR@10 (51.95). Comparable improvements are also observed with both \texttt{Contriever} and \texttt{bge-large-en-v1.5}. These results demonstrate that our approach effectively leverages additional informative contextual signals while suppressing noise, ultimately achieving the strongest performance across all evaluated configurations.

\begin{table}[t!]
\caption{Ablation study on the components of DeCoR. The best performance is highlighted in \textbf{bold}.}
\label{Tab:2}
\resizebox{1.0\columnwidth}{!}{
\begin{tabular}{lcccc}
\toprule
& \textbf{Hits@10}                & \textbf{Hits@4}                & \textbf{MAP@10}                 & \textbf{MRR@10}                 \\ \hline\hline
& \textbf{72.96}         & \textbf{58.51}         & \textbf{23.62}         & \textbf{52.21}         \\ \hline
\rowcolor[HTML]{EFEFEF} (\textit{without}) & & & & \\
\quad (-) \textit{Query Expansion}                                                                             & 68.07                  & 52.45                  & 19.05                  & 42.85                  \\ 
\quad (-) \textit{Query Decomposition}                                                                         & 68.76                  & 55.36                  & 21.65                  & 49.90                  \\
\multirow{2}{*}{\begin{tabular}[c]{@{}l@{}} \quad (-) \textit{Query-aware}\\  \quad\quad\; \textit{Document Compression}\end{tabular}} & \multirow{2}{*}{69.00} & \multirow{2}{*}{56.41} & \multirow{2}{*}{22.16} & \multirow{2}{*}{50.11} \\ &                        &                        &                        &                        \\ 
\quad (-) \textit{concatenate-then-compress}                                                                   & 71.79                  & 58.04                  & 22.26                  & 50.33                  \\ 
\quad (-) \textit{average embedding}                                                                           & 71.79                  & 57.28                  & 21.63                  & 50.63                  \\ \bottomrule
\end{tabular}
}
\end{table}

\subsubsection{Ablation Study}
To assess the contribution of each component of DeCoR to retrieval performance, we conducted an ablation study using 500 queries randomly sampled from the MultiHop-RAG dataset. The results are presented in Table~\ref{Tab:2}. 

The removal of any single component led to a noticeable drop in performance, with the removal of Query Expansion causing the largest degradation (Hits@10: 72.96 → 68.07). This finding underscores that our expansion strategy is fundamental to enhancing retrieval effectiveness. Among the components, the absence of \textit{Query Decomposition} was particularly detrimental, confirming that decomposing complex queries into sub-queries is crucial for covering diverse reasoning steps and expanding semantic diversity, both of which are essential for retrieving multi-layered evidence in multi-hop scenarios. Similarly, excluding \textit{Query-aware Document Compression} also degraded performance, demonstrating the importance of selectively refining context. Rather than indiscriminately adding content, our compression strategy filters and aligns information with the query's intent, thereby mitigating noise and addressing the limitations of naive context expansion.

We also examined two key design choices within DeCoR. First, the \textit{concatenate-then-compress} variant outperforms the document-wise compression variant, which compresses documents individually rather than as an integrated whole. This finding indicates that jointly processing candidate documents enables the model to capture globally salient information, integrate scattered cross-document evidence, and eliminate redundancy, thereby producing a concise and query-focused summary. Second, replacing the \textit{average embedding} strategy with simple concatenation of all text into a single embedding input also degraded performance. This suggests that averaging embeddings yields a more stable and balanced representation, ensuring that the contextual meaning of each component contributes proportionally to the final query embedding.

Overall, the ablation results empirically validate that all components of DeCoR, from Query Decomposition and Query-aware Document Compression to the final embedding strategy, operate synergistically to maximize retrieval performance.

\subsubsection{Impact of Language Models on Performance}

We further analyze the performance of several LLMs such as \texttt{LLaMA-3.1-8B}\cite{dubey2024llama}, \texttt{Mistral-7B}\cite{jiang2023mistral7b}, \texttt{Qwen2.5-7B}\cite{qwen2025qwen25technicalreport}, \texttt{GPT-3.5}\cite{brown2020languagemodelsfewshotlearners}, and \texttt{GPT-4o}\cite{achiam2023gpt}. The effectiveness of the core components of DeCoR, Query Decomposition and Query-aware Document Compression, is closely tied to the capabilities of the underlying LLM. As shown in Table~\ref{Tab:3}, more advanced LLMs achieve higher accuracy in these components, which in turn leads to improved overall retrieval performance. Based on these findings, we selected \texttt{Qwen2.5-7B} for our experiments, as it achieved the best performance among non-commercial models.

This choice provides a noteworthy insight when considered alongside the results in Table~\ref{Tab:1}. Although DeCoR employs the relatively small \texttt{Qwen2.5-7B}, strong baselines such as HyDE and Query2Doc rely on the larger \texttt{GPT-3.5}. Despite this disparity, DeCoR consistently outperforms these baselines, underscoring the resilience of our approach. The observed performance gap may stem from a fundamental methodological distinction. Existing query expansion methods depend heavily on the generative capacity of large-scale LLMs to produce external content, such as pseudo-passages or hypothetical answers. Their effectiveness is intrinsically linked to the scale and raw generation ability of the underlying model.

In contrast, our proposed DeCoR optimizes the retrieval process by restructuring and refining existing information rather than generating new content. It achieves this by decomposing complex queries into simpler sub-questions and selectively distilling relevant evidence from retrieved candidate documents. This design enables DeCoR to fully exploit the reasoning and summarization capabilities of smaller models while reducing reliance on the large-scale generative capacity.

\begin{table}[t!]
\caption{Analysis on LLM Performance. The best performance is highlighted in \textbf{bold}.}
\label{Tab:3}
\resizebox{0.95\columnwidth}{!}{
\begin{tabular}{lcccc}
\toprule
             & \textbf{Hits@10} & \textbf{Hits@4} & \textbf{MAP@10} & \textbf{MRR@10} \\ \hline\hline
\texttt{Llama-3.1-8B-Instruct} & 69               & 54.18           & 21.21           & 48.32           \\ 
\texttt{Mistral-7B-Instruct}   & 68.76            & 54.08           & 21.06           & 47.91           \\ 
\texttt{Qwen2.5-7B-Instruct}   & 71.10            & 57.00           & 22.11           & 50.27           \\ 
\texttt{GPT-3.5-Turbo}      & 72.96            & 58.51           & 22.76           & 51.79           \\ 
\texttt{GPT-4o}       & \textbf{74.76}   & \textbf{59.21}  & \textbf{23.26}  & \textbf{53.46}  \\ \bottomrule
\end{tabular}
}
\end{table}

\section{Conclusion}

In this paper, we address the limitations of existing query expansion methods, particularly their tendency to generate noisy content and their limited effectiveness in complex retrieval scenarios. To overcome these challenges, we propose DeCoR, a framework for structured query expansion in multi-hop retrieval. DeCoR integrates two key components: Query Decomposition, which breaks down a complex query into simpler, independently answerable sub-queries, and Query-aware Document Compression, which distills relevant evidence from retrieved documents. Together, these components contribute to a more balanced and robust query representation. Experiments on the MultiHop-RAG dataset show that DeCoR, despite using a relatively small model (\texttt{Qwen2.5-7B}), consistently outperforms baselines that rely on much larger models such as \texttt{GPT-3.5}. These results demonstrate that, in complex retrieval tasks, strategically enhancing reasoning and summarization capabilities in smaller models can be more efficient and competitive than merely scaling generative capacity. In conclusion, our study highlights that explicitly modeling reasoning steps and selectively refining context are crucial for advancing multi-hop IR. This work potentially represents a meaningful step forward and opens promising directions for future research toward more efficient, effective, and reasoning-aware retrieval systems.

\begin{acks}
This work was supported by the Institute of Information \& Communications Technology Planning \& Evaluation (IITP) grant funded by the Korea government (MSIT) [RS-2021-II211341, Artificial Intelligence Graduate School Program (Chung-Ang
University)] and by the National Research Foundation of Korea (NRF) grant funded by the Korea
government (MSIT) (RS-2025-00556246).
\end{acks}

\section{GenAI Usage Disclosure}
This paper used Generative AI, specifically ChatGPT (OpenAI), to assist with final proofreading and grammatical corrections aimed at improving readability. All AI-suggested edits were thoroughly reviewed and validated by the authors to ensure accuracy, originality, and full compliance with ACM authorship policies.


\bibliographystyle{ACM-Reference-Format}
\balance
\bibliography{sample-base}


\end{document}